\documentclass[preprint,12pt]{elsarticle}
\usepackage{amssymb}
\usepackage{amsmath}
\usepackage{caption}
\usepackage{subcaption}
\usepackage{float}

\journal{Computers \& Fluids}

\begin{document}

\begin{frontmatter}
\title{Interfacial deformation and jetting of a magnetic fluid}
\author[1]{Shahriar Afkhami \corref{cor1}}
\cortext[cor1]{Corresponding author email address: shahriar.afkhami@njit.edu}
\author[1]{Linda J.~Cummings}
\author[2]{and Ian M.~Griffiths}
\address[1]{Department of Mathematical Sciences, New Jersey Institute of Technology, Newark, New Jersey, USA}
\address[2]{Mathematical Institute, University of Oxford, Oxford, England, UK}

\begin{abstract}
An attractive technique for forming and collecting aggregates of magnetic material at a liquid--air interface
by an applied magnetic field gradient was recently addressed theoretically and experimentally [Soft Matter, (9) 2013, 8600-8608]: 
when the magnetic field is weak, the deflection of the liquid--air interface has a steady shape, while for sufficiently strong fields, 
the interface destabilizes and forms a jet that extracts magnetic material. Motivated by this work, we develop a numerical model for the closely related 
problem of solving two-phase Navier--Stokes equations coupled with the static Maxwell equations. We computationally model the 
forces generated by a magnetic field gradient produced by a permanent magnet and so determine the
interfacial deflection of a magnetic fluid (a pure ferrofluid system) and the transition into a jet. 
We analyze the shape of the liquid--air interface during the deformation stage and the critical 
magnet distance for which the static interface transitions into a jet. We draw conclusions on the ability of our numerical model
to predict the large interfacial deformation and the consequent jetting, free of any fitting parameter.
\end{abstract}


\end{frontmatter}
\clearpage

\section{Introduction}

Synthesis and assembly on the nanoscale is an important goal of contemporary science 
and technology. Magnetic nano/microparticles arise in a wide range of industrial and biomedical 
applications, and so are one target for controlled assembly. For example, functionalized magnetic microparticles 
can be used to separate cells \cite{Moldday1977} and magnetic microparticles have been used in microfluidics for cell sorting, blood cleansing, and
magneto-capillary self-assembly (see e.g.~\cite{Stone13} and references therein). When magnetic nanoparticles
such as magnetite are suspended
at high concentration in aqueous or non-aqueous carrier fluids, the entire system behaves as a continuum of
magnetic fluid, known also as a ferrofluid. The rheology and interfacial shape of ferrofluids
can be tuned with external magnetic fields, often in useful ways. An example is the application of ferrofluids in adaptive optics that
has been considered in recent experiments~\cite{Malouin2010,Malouin2011}. The control of ferrofluid properties
using magnetic fields also has applications in mechanical sealing and acoustics \cite{Raj1995}, targeted drug delivery \cite{Yue2012,Cherry2014,Grant2014}
and treatment of retinal detachment \cite{Mefford}.

Thin liquid films and droplets are ubiquitous in nature and also appear in many technological applications. 
The understanding of their dynamical behavior and their stability is therefore of great importance
and has attracted considerable attention in the literature. Recent research into
thin film and droplet flows has resulted in many experimental and theoretical developments, 
including manipulating film flows via external magnetic or electric fields to produce nanoscale patterns.  
In particular, experiments on thin ferrofluid films and droplets have revealed the formation of a wide range of morphologies 
\cite{Baygents98,Stone99,Chen2008,Chen2010,Dickstein1993}. Ferrofluids 
can be manipulated using magnetic forces and have been extensively investigated and widely used in a 
variety of engineering applications; see Rosensweig~\cite{Rosensweig1987} and a more recent review by Nguyen \cite{Nguyen2011}. 
Normal field instability of ferrofluid films (and the equivalent electric field problem) have been extensively studied in the past, see e.g.~\cite{Craster2005,SAK14}.
However, despite the increase in the number of applications, surprisingly, little can be found in the 
literature on the direct numerical simulations of thin ferrofluid films in the presence of a nonuniform magnetic field (such as is
produced by a spherical magnet) and therefore our understanding of the instabilities that may occur in these flows is limited.

\begin{figure}[t]
\begin{center}
 \includegraphics[scale=0.5, trim=25mm 20mm 10mm 10mm, clip=true]{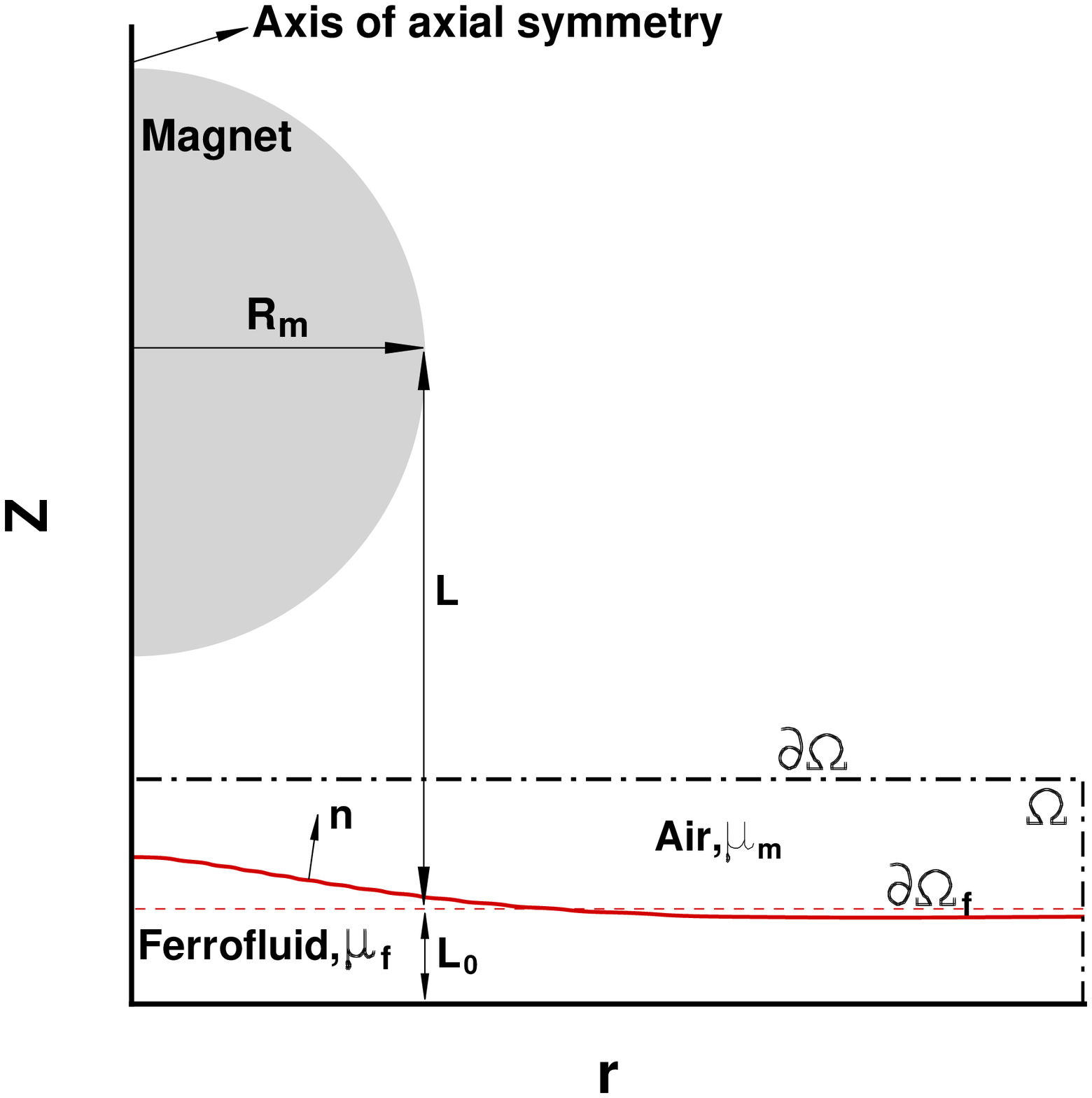}
\end{center} 
\caption{A schematic illustrating the computational setup and the coordinate system used. A spherical permanent magnet of radius $R_m$
is centered at distance $L$ from the initially undeformed film (red dashed line), which has a depth
$L_0$. The magnetic force deforms the interface, $\partial \Omega_f$, into a hump (red solid line).
A unit normal outwards from the interface is denoted by ${\bf n}$.  A typical computational domain, 
$\Omega$, and its boundary, $\partial \Omega$, is shown by the dash-dotted line.
Due to symmetry, only half of the domain is simulated.}
\label{fig:initial}
\end{figure}

An attractive technique for forming and collecting aggregates of magnetic material at a liquid--air interface by 
an applied magnetic field was recently addressed experimentally and theoretically by Tsai {\it et al.} \cite{Griffiths13}. 
In the experiments described in \cite{Griffiths13}, a water-based ferrofluid (EMG805, Ferrotec), 
with a density of $1200$ kg m$^{-3}$ and viscosity of $3$ mPa s, is suspended in a shallow reservoir containing 
deionized water, with a density of $1000$ kg m$^{-3}$ and viscosity of $1$ mPa s, to form the magnetic mixture.
A spherical permanent magnet is then slowly brought close to the magnetic mixture allowing the ferrofluid 
to aggregate and form a static hump at the liquid--air interface (cf. Fig.~\ref{fig:initial}). In these experiments, a distinct
boundary that separates the magnetic and non-magnetic regions is observed. When the magnet is held
sufficiently close to the liquid--air interface, the hump destabilizes and transforms to a jet.
The theoretical approach developed in \cite{Griffiths13} describes a steady-state mathematical model for 
the behavior of the magnetic-particle-laden fluid and the particle-free fluid regions.
The mathematical model results in \cite{Griffiths13} show excellent agreement with the experimental data.

Motivated by this work, here we develop a numerical model for a closely related problem: 
we computationally model the magnetically induced interfacial deflection of a magnetic fluid (ferrofluid)
and the transition into a jet by a magnetic field gradient from a permanent magnet placed above the free surface.
The system we study differs from that considered by Tsai {\it et al.} \cite{Griffiths13}: we consider a pure ferrofluid system, while Tsai {\it et al.} model a system with both magnetically dominated
and non-magnetic regions. Figure \ref{fig:initial} shows a
schematic illustrating the setup we consider: the magnetic region occupied by pure ferrofluid, the liquid--air interface, and 
the spherical permanent magnet. The deformation of the ferrofluid--air interface arises as a result of the magnetic field gradient induced by the spherical permanent magnet held above the fluid. 
We solve the two-phase Navier--Stokes equations coupled with the static Maxwell equations. 
We analyze the shape of the liquid--air interface during the deformation stage and the critical
magnet distance (from the undeformed free surface), for which the static interface transitions into a jet. We draw conclusions regarding the ability of our numerical
model to predict the large interfacial deformation and the consequent jetting, free of any fitting parameters.
The numerical model provides a realistic and accurate framework for predicting  
the evolution of magnetic liquids based on the Navier--Stokes equations. 

We describe the details of the numerical model in Sec.~\ref{sec:model}.
In Sec.~\ref{sec:methods}, we describe a numerical boundary condition that may be implemented to simulate non-uniform magnetic fields.
In Sec.~\ref{sec:results}, we present the numerical results and the comparison with experimental observations.
In Sec.~\ref{sec:conc}, we give an overview and future outlook for improving our modeling.   

\section{Mathematical Model}
\label{sec:model}
Here we briefly describe the theoretical models that serve as a basis for the proposed
numerical studies. The coupled motion of a ferrofluid surrounded by a non-magnetic fluid is 
governed by the (static) Maxwell equations, the Navier--Stokes equations, and a 
constitutive relationship for the  magnetic induction ${\bf B} $, 
magnetic field  ${\bf H}$, and the magnetization ${\bf M}$.  The magnetostatic
Maxwell equations for a non-conducting ferrofluid are, in SI units,
\begin{eqnarray*}\label{eq:main_eq}
\nabla\cdot{\bf B}&=&0,\quad \nabla\times{\bf H}=0,\quad
{\bf B}({\bf x},t) =  \left\{ 
\begin{array}{ll}
                \mu_{f} {\bf H }  &  \mbox{in ferrofluid} \\
                \mu_m {\bf H}   &  \mbox{in matrix,}
\end{array}        
\right. 
\label{classical}
\end{eqnarray*}
where $\mu_f$ denotes the magnetic permeability of the ferrofluid and $\mu_m$ is the
permeability of the matrix fluid.  For our application, the matrix fluid is air, which
has a permeability very close to that for a  vacuum, $\mu_o$. Therefore, we shall consider  
$\mu_m=\mu_o$ throughout this mathematical model. A magnetic scalar potential 
$\psi$ is defined by ${\bf H}=\nabla\psi$, and satisfies
\begin{equation}
\nabla\cdot(\mu\nabla\psi)=0,  \label{eq:laplace}
\end{equation}  
where $\mu=\mu_o$ and $\mu_{f}$ in the matrix and ferrofluid, respectively. We will assume that the magnetization is a linear function of the
magnetic field given by ${\bf M} = \chi {\bf H}$, where $\chi= (\mu_f/\mu_o - 1)$ is the 
magnetic susceptibility \cite{ATRRPWR}. The magnetic induction ${\bf B}$ is therefore  
${\bf B}=\mu_o({\bf H}+{\bf M})=\mu_o(1+\chi){\bf H}$.

The fluid equation of motion is described by the conservation of mass and momentum (Navier--Stokes) equations
\begin{eqnarray}  
\nabla \cdot {\bf u}&=&0,\\
\rho \left( \frac{\partial{\bf u}}{\partial t} + ({\bf u} \cdot \nabla){\bf u} \right)&=&-\nabla p + \nabla\cdot \bigl(2\eta {\bf D} \bigr) + \nabla\cdot\tau_m + {\bf F}_s + \rho {\bf g},
\label{eq:motion}
\end{eqnarray}      
where ${\bf F}_s$ denotes the surface tension force per unit volume (presented as a body force~\cite{Brackbill92}), $p$ is pressure, ${\bf u}$ is velocity,
${\bf D}=\frac{1}{2}\big( \nabla{\bf u} +(\nabla{\bf u})^T\big)$ 
is the rate of deformation tensor (where $T$ denotes the transpose),
$\eta$ is viscosity, $\rho$ is density, $\tau_m$ is the magnetic 
stress tensor, and ${\bf g}$ is the gravitational acceleration. The total stress is $\tau =-p{\bf I} +2\eta{\bf D}+\tau_m$, where ${\bf I}$
denotes the identity operator.
The magnetic stress tensor of an incompressible, isothermal, magnetizable medium is \cite{Rosensweig}
\begin{eqnarray} 
\tau_m  =  - \frac{\mu_o}{2} H^2{\bf I} + \mu {\bf H}{\bf H}^T,\nonumber
\end{eqnarray}
where $H=|{\bf H}|$.  These equations must be solved subject to suitable boundary and initial conditions, discussed in Sec.~\ref{sec:methods} below.

\section{Numerical Methodologies}
\label{sec:methods}
We will use an Eulerian framework, where the material moves through a stationary mesh, 
and therefore a special procedure will be required to track the interface between fluids.
We will use the volume of fluid (VOF) method to track the interface
between the ferrofluid and the matrix fluid \cite{FCDKSW2006,AB2008,AB2009}.
In this way, the VOF formulation describes each fluid by assigning a volume fraction function, $f(r,z,t)$, as
\begin{eqnarray}
  f(r,z,t) & = & \left\{ \begin{array}{ll}
        1 & \mbox{in ferrofluid} \\
        0 & \mbox{in matrix} \end{array} \right.
\end{eqnarray}
(see Fig.~\ref{fig:initial} for the polar coordinates $(r,z)$ used here).
The position of the interface is computed from $f(r,z,t)$
by reconstructing the curve where the step discontinuity takes place. 
In this work, the reconstruction is a `piecewise linear interface calculation' (PLIC) \cite{LR98}.
To track the interface, $f(r,z,t)$ is advected by the flow
\begin{equation}\label{eq:VOF}
\frac{\partial f}{\partial t} + \nabla \cdot ({\bf u}f) = 0. 
\end{equation}
In the Navier--Stokes equations (Eq.~(\ref{eq:motion})),
${\bf F}_s$ includes the continuum body force due to interfacial tension
\begin{equation}
{\bf F}_s=\sigma \kappa  \delta_S {\bf n}, 
\label{eq:ST}
\end{equation}
where the unit normal ${\bf n} =\nabla f/|\nabla f|$ and $\delta_S=|\nabla f|$ is the
delta function at the interface. 


\subsection{An Optimal Method for Simulating Non-Uniform Magnetic Fields}  
\label{sec:BC}
Special care must be taken when computing the
solution of the Maxwell equations to account, accurately and robustly, for the non-uniformity of the magnetic field.
The boundary condition on the magnetic field is reconstructed from an existing solution of the magnetic field due 
to a spherical magnet, in the absence of the ferrofluid.
We let $\psi_a$ denote the magnetic potential for this infinite-domain analytical solution,
satisfying Laplace's equation, in cylindrical coordinates with azimuthal symmetry, 
$$\frac{1}{r}\frac{\partial}{\partial r}\left(r\frac{\partial\psi_a}{\partial r} \right)+\frac 
{\partial^2 \psi_a} {\partial z^2}=0.$$
We assume a magnetic potential generated by a uniformly magnetized spherical magnet, as in \cite{Griffiths13}, hence
\begin{eqnarray}
\psi_a(r,z)= \left\{\frac{M_mR^3_m(L-(z-L_0))r}{3\left[(L-(z-L_0))^2+r^2\right]^{3/2}}\right\} + \mbox{constant},
\label{eq:analytical}
\end{eqnarray}
where $M_m$ is the magnetization of the magnet and $R_m$ is the radius of the magnet.
The magnet center is at $(r,z)=(0,L+L_0)$, where $L$ is 
the magnet distance from the undeformed interface and $L_0$ is the initial unperturbed interfacial thickness
in our problem setup (see Fig.~\ref{fig:initial}).
The analytical magnetic field ${\bf H}_a=\nabla\psi_a$ is thus 
\begin{eqnarray}
{\bf H}_a= \left\{\frac{M_mR^3_m((z-L_0)-L)r}{\left[(L-(z-L_0))^2+r^2\right]^{5/2}}\right\}{\hat {\bf r}} 
+ \left\{\frac{M_mR^3_m\left[2((z-L_0)-L)^2-r^2\right]}{3\left[(L-(z-L_0))^2+r^2\right]^{5/2}}\right\}{\hat {\bf z}},
\label{eq:H}
\end{eqnarray}
where $\hat {\bf r}$ and $\hat {\bf z}$ denote the unit vectors in the $r$ and $z$ directions, respectively.
This analytical solution, Eq.~(\ref{eq:analytical}), provides the boundary conditions that we will impose on the magnetic potential at the boundaries of 
our computational domain.

The numerical magnetic potential $\psi$ is calculated by solving Eq.~(\ref{eq:laplace}) on the computational
domain $\Omega$ (see Fig.~\ref{fig:initial}) employing the boundary conditions obtained using the analytical solution $\psi_a(r,z)$ above.
In all simulations presented, the spherical magnet lies outside the computational domain.
The numerical solution to the elliptic
partial differential equation (Eq.~(\ref{eq:laplace})) will be complicated by the fact that the magnetic permeability, $\mu$,
experiences a jump across the interface \cite{ATRRPWR}. We will 
use a multigrid algorithm, as described in \cite{ATRRPWR}, to solve Eq.~(\ref{eq:laplace})
subjected to the boundary conditions described below. 

The boundary conditions for $\psi$ 
on the domain boundaries $\partial \Omega$ are defined as
\begin{eqnarray}
\frac{\partial \psi}{\partial n} = \frac{\partial \psi_a}{\partial n} & \mbox{on $\partial \Omega$},
\end{eqnarray}
where $\partial /\partial n={\bf n}_\Gamma\cdot\nabla$, and ${\bf n}_\Gamma$  denotes the normal to the 
boundary $\partial \Omega$.
In order to impose the boundary condition in our numerical model,
we perform a transformation of variables to $\zeta$: $\psi=\psi_a+\zeta$, where
$\psi_a$ is the potential field without the magnetic medium, Eq.~(\ref{eq:analytical}).
One can then rewrite Eq.~(\ref{eq:laplace}) such that
\begin{equation}
\nabla\cdot(\mu\nabla\zeta)=-\nabla\cdot(\mu\nabla\psi_a),\label{magnet2}
\end{equation}
where $\nabla\cdot(\mu\nabla\psi_a)$ vanishes everywhere except on the surface
between the ferrofluid/matrix interface $\partial \Omega_f$ and 
\begin{eqnarray}
\frac{\partial \zeta}{\partial n} = 0 & \mbox{on $\partial \Omega$}.\label{BCmagnet2}
\end{eqnarray}

Though we do not use this in the following analysis, we note that our numerical approach may also be used to model the nonlinear magnetic susceptibility of ferrofluids;
for example, the well-known Langevin function
$L(\alpha)$=$\coth \alpha-\alpha^{-1}$ can be implemented to describe the magnetization 
behavior of the ferrofluid versus the strength of the magnetic field ${\bf H}$,
\begin{eqnarray}
{\bf M} ({\bf H}) = M_s L\left(\frac{\mu_o m |{\bf H}|}{k_B T}\right)\frac{{\bf H}}{|{\bf 
H}|}, \label{langevin1}
\end{eqnarray} 
where the saturation magnetization $M_s$ and the magnetic moment of the particle
enter as parameters, $T$ denotes the absolute temperature, and $k_B$ is the Boltzmann constant.

\subsection{Pressure, Velocity, and Volume Fraction Boundary Conditions}
We impose symmetry boundary conditions at $r=0$.
At the top, bottom and right boundaries, we impose solid wall boundary conditions for the pressure
and the velocities. The boundary condition for the volume fraction
function at the top and bottom walls is that $f=0$ and for the right wall is that
the interface has zero slope $\frac{\partial f}{\partial r}=0$.

\section{Results and Discussion}
\label{sec:results}

Numerical simulations are presented in three parts. First we present tests of the
numerical implementation by focusing on the resulting magnetic fields and by studying
the convergence of the solution for the magnetic field with grid refinement. Second, we
present several numerical solutions of our system that demonstrate both the steady deflection
regime, and the transition from this regime to the unstable jetting regime.
A dimensional analysis in \cite{Griffiths13} shows that the steady interfacial deformation
can be characterized by the magnetic Bond number that captures the dominant balance between
the magnetic and the surface tension forces.
Following this approach, we present a parametric study by varying the magnetic Bond number within 
the steady interface deflection regime.

Our numerical setup closely follows that in \cite{Griffiths13}; a spherical permanent magnet 
of radius $R_m$ and magnetization $M_m=10^6$ A m$^{-1}$ is placed at a distance $L$ 
from the initially undeformed ferrofluid film (see Fig.~\ref{fig:initial}).
The ferrofluid has a density $\rho_f=1200$ kg m$^{-3}$, viscosity $\eta_f=3$ mPa s, and
surface tension $\sigma=0.07$ N m$^{-1}$. The ferrofluid is assumed to have a constant magnetic susceptibility, $\chi$,
which will be determined by comparing the numerical result with the experimental data in \cite{Griffiths13} (note that
this value is often not reported by manufacturers so is determined by users).

\subsection{Magnetic Field and Imposed Numerical Boundary Condition}
\begin{figure}[t]
\begin{center}
 \includegraphics[scale=0.5, trim=30mm 20mm 20mm 0mm]{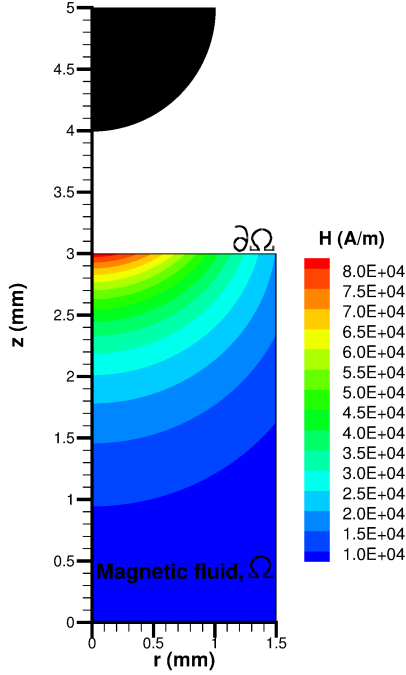}
\end{center} 
\caption{The computational domain, $\Omega$, and the computed magnetic field strength, $H$, with a drawing of a permanent magnet overlaid.
         The spherical magnet lies outside the computational domain. 
         Number of grid points in the $z$-direction is $N_z=240$ and in the $r$-direction is $N_r=120$.}
\label{fig:mfield}
\end{figure}
\begin{figure}[th]
\begin{center}
 \includegraphics[scale=0.65]{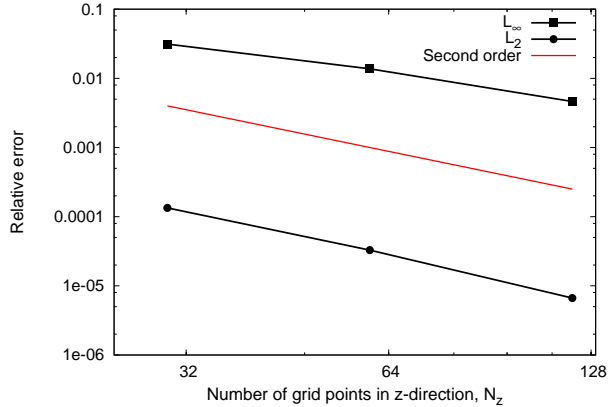}
\end{center} 
\caption{The convergence of the $L_2$ and $L_{\infty}$ relative error norms between the $H(r,z)$
computed at $\Delta=L_z/N_z(=L_r/N_r)$ and at $\Delta_{\mbox{max}}=L_z/240$, for the computational setup illustrated in Fig.~\ref{fig:mfield}.
}
\label{fig:convergence}
\end{figure}

We first demonstrate the effectiveness of our methodology for computing the magnetic field 
by presenting the results of the simulated applied magnetic field using the numerical boundary conditions
described in Sec.~\ref{sec:BC}. The following relative error norms are defined 
$$
L_2(H,\Delta)=\frac{||H||_{2,\Delta}-||H||_{2,\Delta_{\mbox{max}}}}{||H||_{2,\Delta_{\mbox{max}}}},\quad
L_{\infty}(H,\Delta)=\frac{||H||_{\infty,\Delta}-||H||_{\infty,\Delta_{\mbox{max}}}}{||H||_{\infty,\Delta_{\mbox{max}}}},
$$
where
$$||H||_{2,\Delta}=\sqrt{\Sigma_i \Sigma_j H^2_{ij}\Delta^2} \quad\mbox{and}\quad ||H||_{\infty,\Delta}={\mbox{Max}}\left(|H_{ij}|\right)_{\Delta},$$
where $i$ and $j$ are indices of a computational cell.
Figure \ref{fig:mfield} shows the computational domain $\Omega$: $r \in (0,L_r)$ and $z \in (0,L_z)$, where $(L_r, L_z)=(1.5$ mm, $3$ mm$)$, with a $1$ mm radius
permanent magnet placed $5$ mm above the $r$-axis. The mesh size $\Delta=L_r/N_r=L_z/N_z$, where $N_r$ and $N_z$ are the number of grid points in
$r$ and $z$ directions, respectively. The computed magnetic field strength, $H$, is also shown in Fig.~\ref{fig:mfield}.

Figure \ref{fig:convergence} exhibits the convergence of the numerical method with spatial resolution for computing the magnetic field.
As illustrated, a second-order convergence is obtained for the mesh refinement for both $L_2$ and $L_{\infty}$ error  measures. The figure also reveals the smallness of the errors even at a coarse grid resolution. Finally the accuracy of the numerical solution is also assessed by comparing the computed 
magnetic field with the exact solution for a spherical magnet in an unbounded region given by Eq.~(\ref{eq:H}); the maximum errors in $H$ are of the order $0.01\%$.
These results confirm the overall consistency and accuracy of the computational scheme with the implemented numerical boundary conditions
as described in Sec.~\ref{sec:BC}.

\subsection{Interfacial Deflection and Transition to Jetting}

Tsai {\it et al.} describe the results of experiments and mathematical analysis of the
deformation of a free surface by a magnetic force from a spherical permanent magnet \cite{Griffiths13}. 
As mentioned above, in that work the system consists of both magnetic and non-magnetic
regions such that the magnetic particles collect at the interface, causing deformation 
of the free surface to form a hump; and when the magnet is brought sufficiently close, the hump transitions to a jet.
Within the non-magnetic region, the pressure drop across the static interface, and hydrostatic stresses, are accounted for by
the Young--Laplace equation; within the magnetic region, the liquid is treated as a static ferrofluid 
that has a homogeneous magnetic susceptibility, and the modified form of the Young--Laplace equation (modified to account for the fluid magnetization)
describes the pressure drop across the static interface. The steady state of the interface shape is then determined by two second-order differential equations
solved by imposing appropriate boundary conditions. By exploiting the smallness of the vertical deflection of
the fluid surface compared to the horizontal extent occupied by the
paramagnetic particles, and the smallness of the magnetic susceptibility, Tsai {\it et al.} also
analyze their mathematical model asymptotically in the non-jetting regime. Using the asymptotic
solution and a scaling argument, they compute the maximum interfacial deflection at the center of the hump
to predict when the transition to jetting occurs \cite{Griffiths13}.

\begin{figure}[H]
\begin{center}
 \includegraphics[scale=0.85, trim=0 15mm 0 10mm, clip=true]{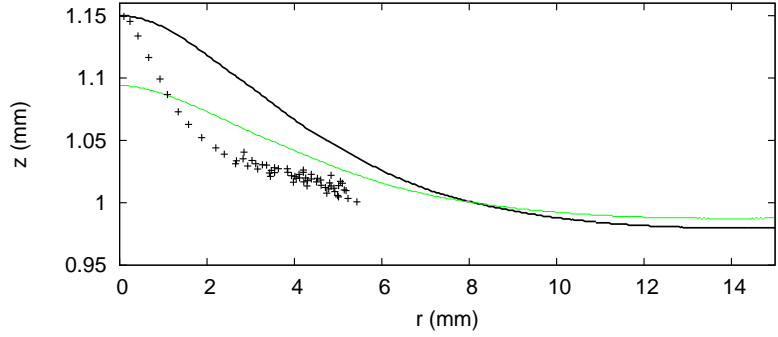}\\
 (a)\\
 \includegraphics[scale=0.85, trim=0 15mm 0 10mm, clip=true]{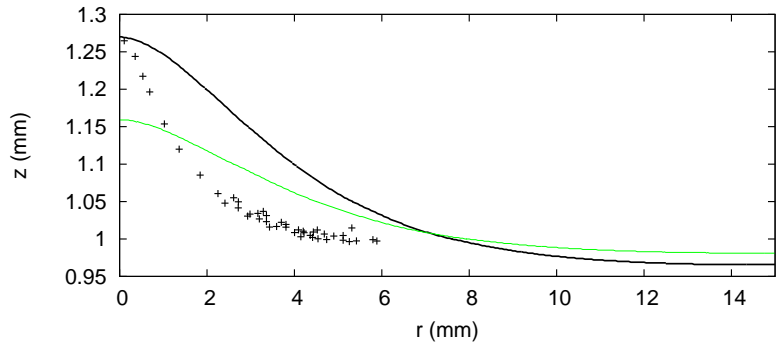}\\
 (b)
\end{center} 
\caption{Experimental data (not pure ferrofluid) from \cite{Griffiths13} shown alongside our computed steady-state ferrofluid--air interfacial profiles for (a) $L=6.33$ mm and (b) $L=5.65$ mm.  Black solid lines have $\chi=7.5\times 10^{-4}$, chosen to match maximum free surface height to data; green solid lines have $\chi=5\times 10^{-4}$, chosen to fit the data as a whole. The results illustrate the height of the hump and the radial spread of the ferrofluid--air interface.
       }
\label{fig:comparison}
\end{figure}

The system we investigate computationally here differs from that of \cite{Griffiths13} because
our liquid region contains a homogeneous ferrofluid. Despite this difference, here we also observe
the same qualitative features:
when the magnet is positioned sufficiently far from the ferrofluid, the deformation of the
interface is static; and when we reduce the distance between the magnet and the undeformed interface, 
the deformation destabilizes into jet formation. 
In Fig.~\ref{fig:comparison}, we present the results
of direct computations of our system in the static deformation regime. In these simulations, the computational domain is $L_r=15$ mm by $L_z=2.5$ mm and
the undeformed film is initially located at $L_0=1$ mm. A magnet of radius $R_m=3.2$ mm is placed at 
distance $L=6.33$ mm  (Fig.~\ref{fig:comparison}a) and $5.65$ mm (Fig.~\ref{fig:comparison}b) from the undisturbed film. 
In Fig.~\ref{fig:comparison}, we also reproduce the experimental results of \cite{Griffiths13} to compare and contrast the two systems.
We use the comparison 
of the computed steady-state hump height, $h$, with the experimental measurement to determine the magnetic susceptibility of the ferrofluid. This comparison gives
$\chi=7.5\times 10^{-4}$, a value consistent with the finding in \cite{Griffiths13}. 
In Fig.~\ref{fig:comparison}, we also demonstrate the results when using $\chi=5\times 10^{-4}$, chosen to fit the data as a whole
rather than just fitting to the maximum hump height.
Although the results demonstrate consistency between the direct numerical solutions
and the experiments in capturing the height 
of the hump, the agreement in the profile is not quantitative, owing to fact that the experimental system is not pure ferrofluid.

\begin{figure}[t]
\begin{center}
 \includegraphics[scale=0.85, trim=0 15mm 0 10mm, clip=true]{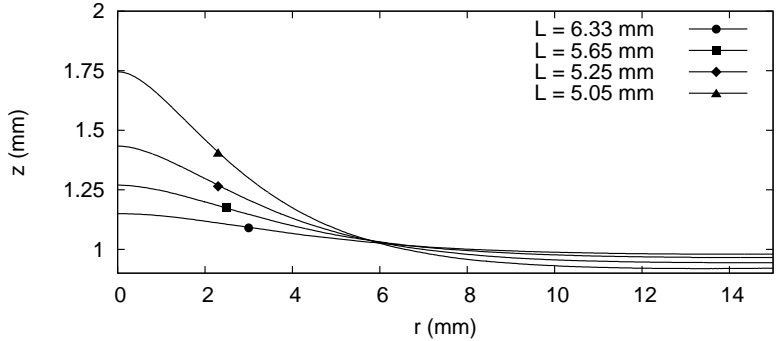}
\end{center} 
\caption{Static ferrofluid--air interfaces when varying $L$; for $L\lesssim 5.05$ mm, no static hump can form.
         A decrease in $L$ results in decreasing the radial spreading. The hump also becomes less pronounced
         with increasing $L$.}
\label{fig:transition}
\end{figure}

\begin{figure}[t]
\begin{center}
 \includegraphics[scale=0.65]{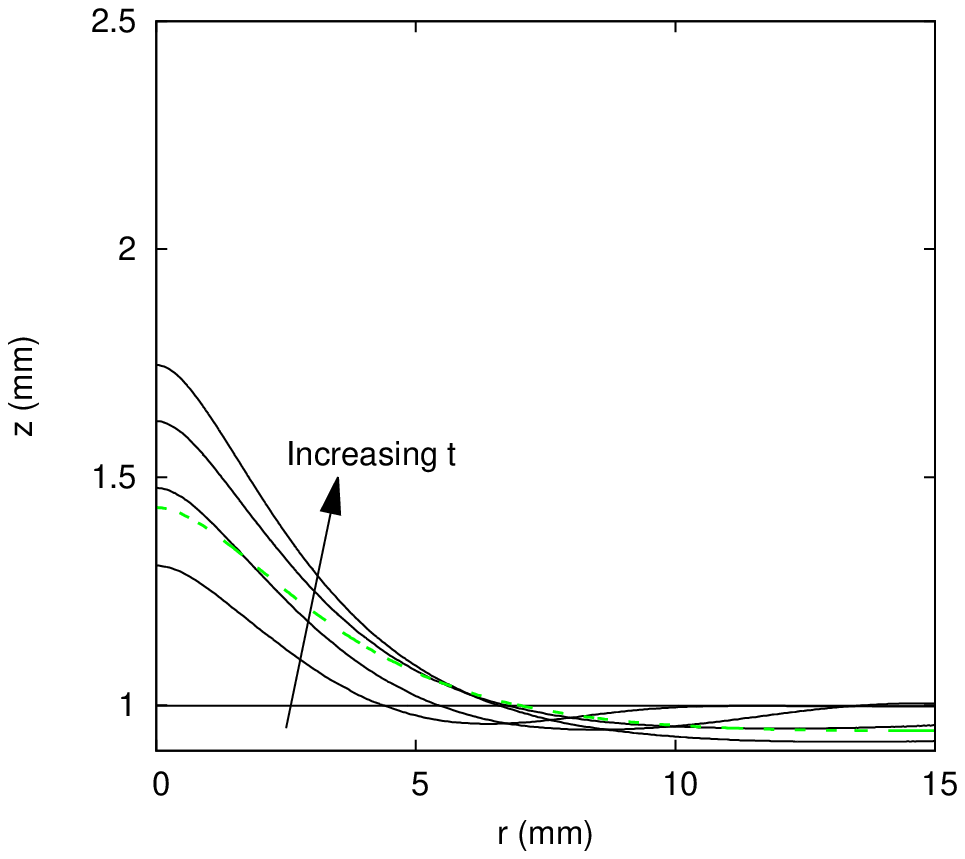}
 \includegraphics[scale=0.65]{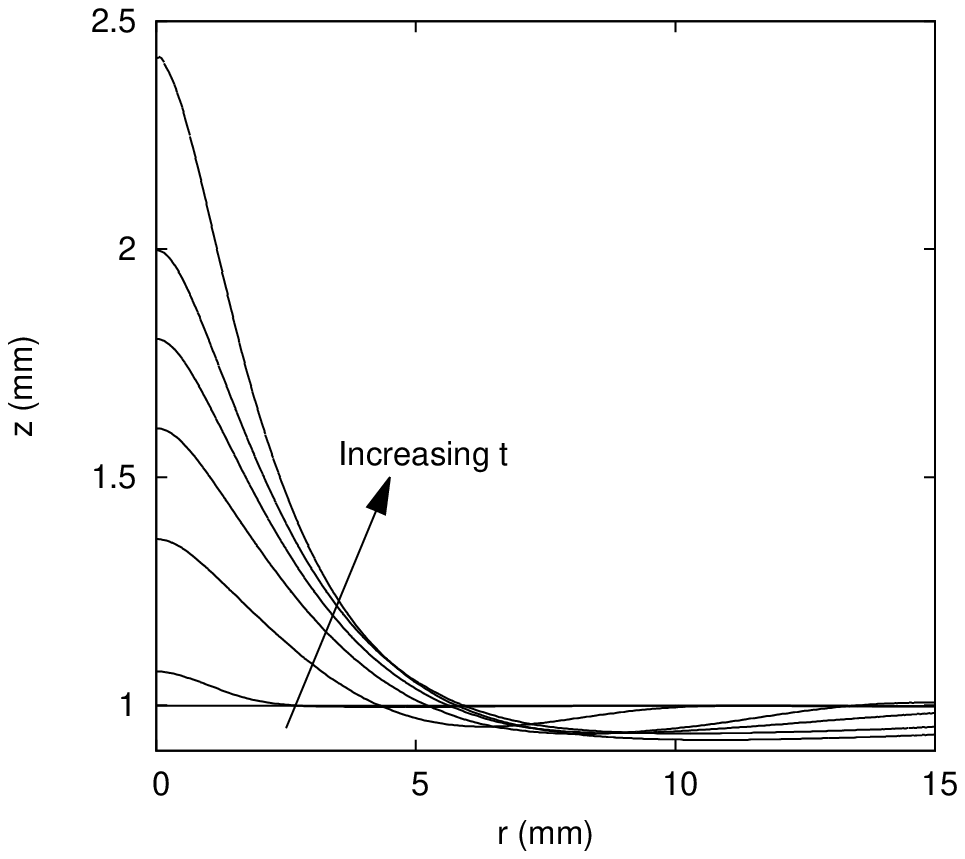}\\
\end{center} 
\hspace{35mm} (a) \hspace{60mm} (b)
\caption{The ferrofluid--air interface. (a) $t=0,25,50,100,200,450$ ms (from bottom to top) for $L=5.25$ mm. 
         The interface deforms and reaches a static state (green dashed line); the interface undergoes a transient overshoot before settling down to a static shape.
         (b) $t=0,5,25,50,100,150,200,248$ ms (from bottom to top) for $L=4.95$ mm. No sustainable steady interfacial deflection can be obtained and
         the dynamic interface grows until it touches the top boundary of the computational domain at $250$ ms (not shown here). The magnet radius $R_m=3.2$ mm and
         the experimentally fitted value of the magnetic susceptibility $\chi=7.5\times 10^{-4}$.}
\label{fig:time}
\end{figure}

With this caveat in mind, we next investigate the transition to jetting with $\chi = 7.5\times 10^{-4}$. In Fig.~\ref{fig:transition},
we show a series of static ferrofluid--air interface profiles as the distance $L$ between magnet center and undeformed ferrofluid interface 
is reduced. 
As the magnet separation $L$ is decreased incrementally, we find that there exists a threshold for the jetting transition:
for $L\lesssim 5.05$ mm, 
we no longer see the free surface evolving to a static configuration; instead, the ferrofluid--air interface destabilizes and forms a jet.
The maximum sustainable steady interfacial deflection that we were able to compute occurs at 
$L=5.05$ mm, consistent with the experimental finding in \cite{Griffiths13}. Also consistent with the experimental 
observations, we find that the steady interfacial deflection increases and the radial spreading decreases, with decreasing $L$, until 
the hump transitions to jetting. 
In Fig.~\ref{fig:time}, we plot interface evolution when the magnet is sufficiently far away ($L=5.25$ mm) that the interface reaches a static state (Fig.~\ref{fig:time}(a));
and when it is close enough ($L=4.95$ mm) that jetting occurs (Fig.~\ref{fig:time}(b)),
where no sustainable static interfacial deflection can be obtained. 
Interestingly, for $L=5.25$ mm (Fig.~\ref{fig:time}(a)), we find an overshoot in deformation; i.e., an initial elongation followed by a retraction to a static state.
We attribute the overshoot to the competition between magnetic, capillary, and inertial effects.
Also, as illustrated, for $L=4.95$ mm (Fig.~\ref{fig:time}(b)), the interface becomes unstable and stretches until it touches
the top boundary of the computational domain while for $L=5.25$ mm, a stable interface is achieved. 
We also note that, when the interface destabilizes, the deflection increases very rapidly as the ferrofluid--air interface approaches the magnet. 

\begin{figure}[t]
\begin{center}
 \includegraphics[scale=0.75]{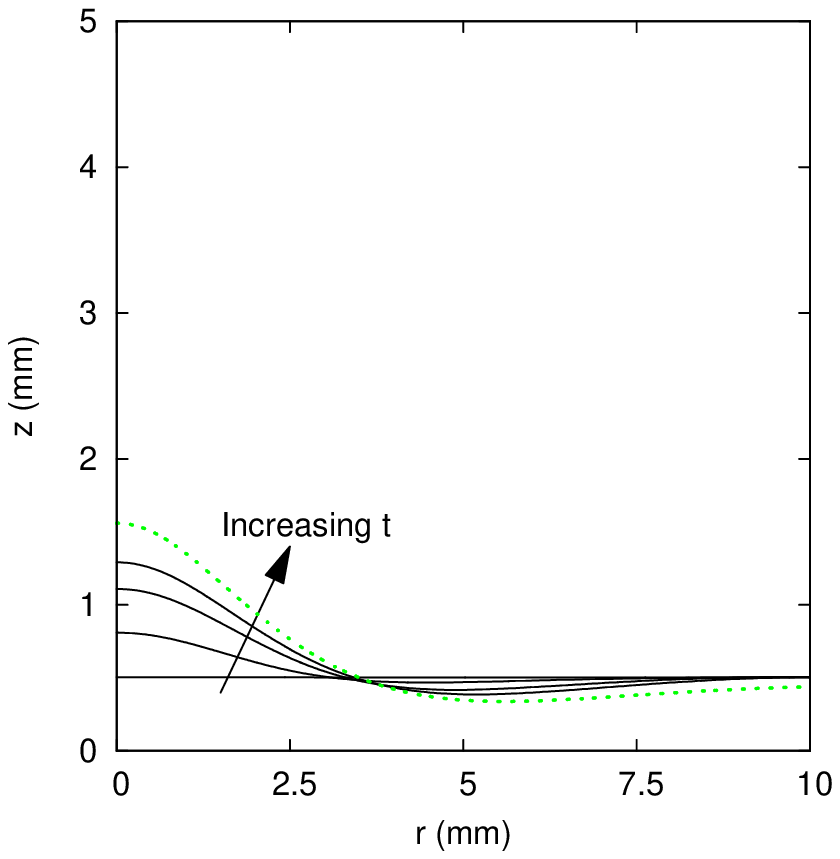}
 \includegraphics[scale=0.75]{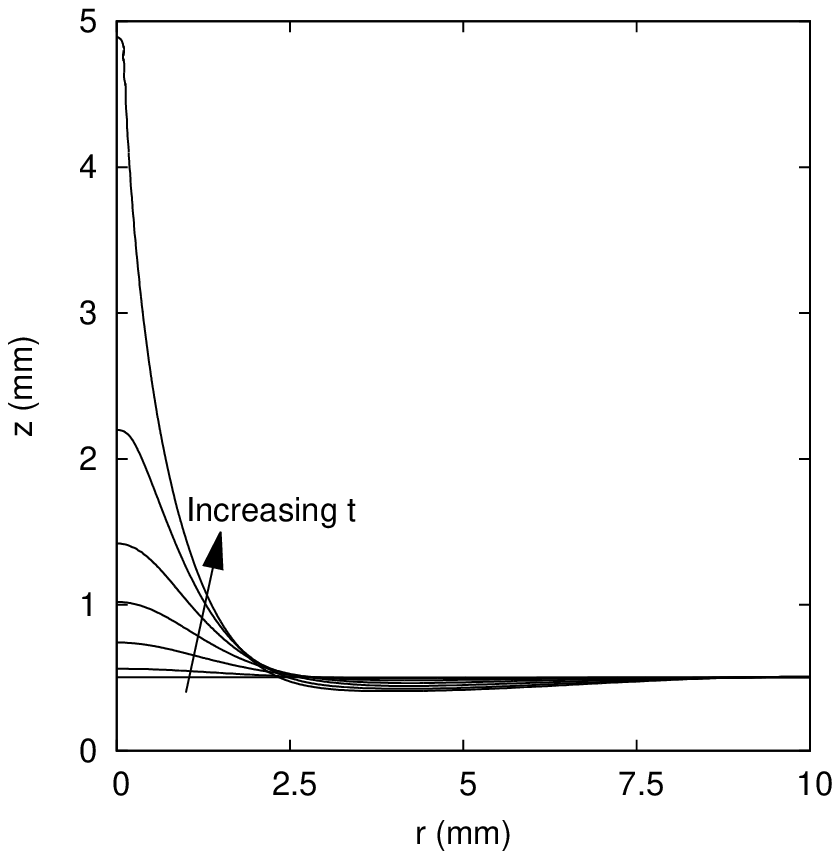} 
\end{center} 
 \hspace{35mm} (a) \hspace{60mm} (b)
\caption{The ferrofluid--air interface. (a) $t=0,12.5,25,37.5$ ms (from bottom to top) for $\chi=0.75$.
The interface deforms and reaches a static state (green dashed line) at around $t=150$ ms.
(b) $t=0,2.5,5,7.5,10,12.5,15$ ms (from bottom to top) for $\chi=1.75$.
The interface touches the top boundary of the computational domain at around $16$ ms (not shown here). 
$L=6.75$ mm and $R_m=1.5$ mm. The flat interface shows the initial position of the ferrofluid--air interface.}
\label{fig:xi1p75-L6p5}
\end{figure}

\begin{figure}[t]
\begin{center}
 \includegraphics[scale=0.85]{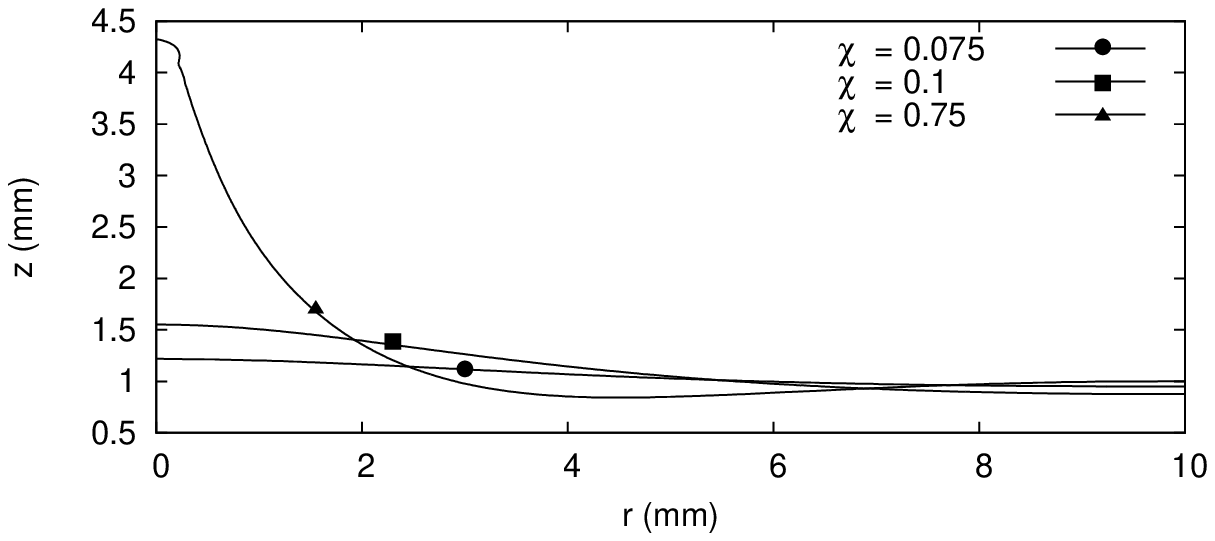}
\end{center} 
\caption{The ferrofluid--air interface when varying $\chi$; for $\chi \gtrsim 0.1$, no static hump can form. 
Increasing $\chi$ results in transition to jetting. 
For $\chi=0.75$, the dynamic interface stretches until it touches the top boundary of the computational domain, while
the interface reaches a static state for $\chi=0.075$ and $\chi=0.1$. In all cases $L=6$ mm and $R_m=1.5$ mm.
A secondary instability (or necking that will grow into a pinch-off structure) appears to develop at the tip of the jet for $\chi=0.75$.}
\label{fig:chi}
\end{figure}

We also investigate the effect of increasing the magnetic susceptibility and the magnet size. 
In Fig.~\ref{fig:xi1p75-L6p5}(a), we plot the ferrofluid--air interface for $\chi=1.75$ as a function of time for $L=6.75$ mm and $R_m=1.5$ mm.
We also plot the case when $\chi=0.75$ in Fig.~\ref{fig:xi1p75-L6p5}(b). These results indicate that increasing the magnetic susceptibility can 
destabilize a previously stable interface, moving it into the jetting regime; and that when the magnet size
is reduced, a much higher magnetic susceptibility is required to destabilize the interface into jetting. These predictions
can be used to find the optimal magnet size for destabilizing the interface based on the susceptibility of the magnetic fluid. For a fixed magnet 
distance $L$ and magnet size $R_m$, transition to jetting occurs for a certain value of the magnetic susceptibility $\chi$.
When the magnet is brought slightly closer (decrease $L$), the transition occurs for a smaller $\chi$ value. Figure \ref{fig:chi}
shows the results for increasing the magnetic susceptibility, which leads to an increase in the 
interfacial deflection for $L=6$ mm and $R_m=1.5$ mm. This figure shows that, for a sufficiently high magnetic susceptibility, the
surface tension of the ferrofluid can no longer sustain the deformation and jetting again occurs.
We find that in this case, for $\chi \gtrsim 0.1$, no static hump can form. 
We also note that for $\chi = 0.75$ in Fig.~\ref{fig:chi},
a secondary instability appears to form at the tip of the jet, although this instability will not have sufficient time to grow
(most likely into a pinch-off structure) before the tip touches the magnet.

\subsection{Scaling Model for Interfacial Deflection}
\begin{figure}[t]
\begin{center}
 \includegraphics[scale=0.85]{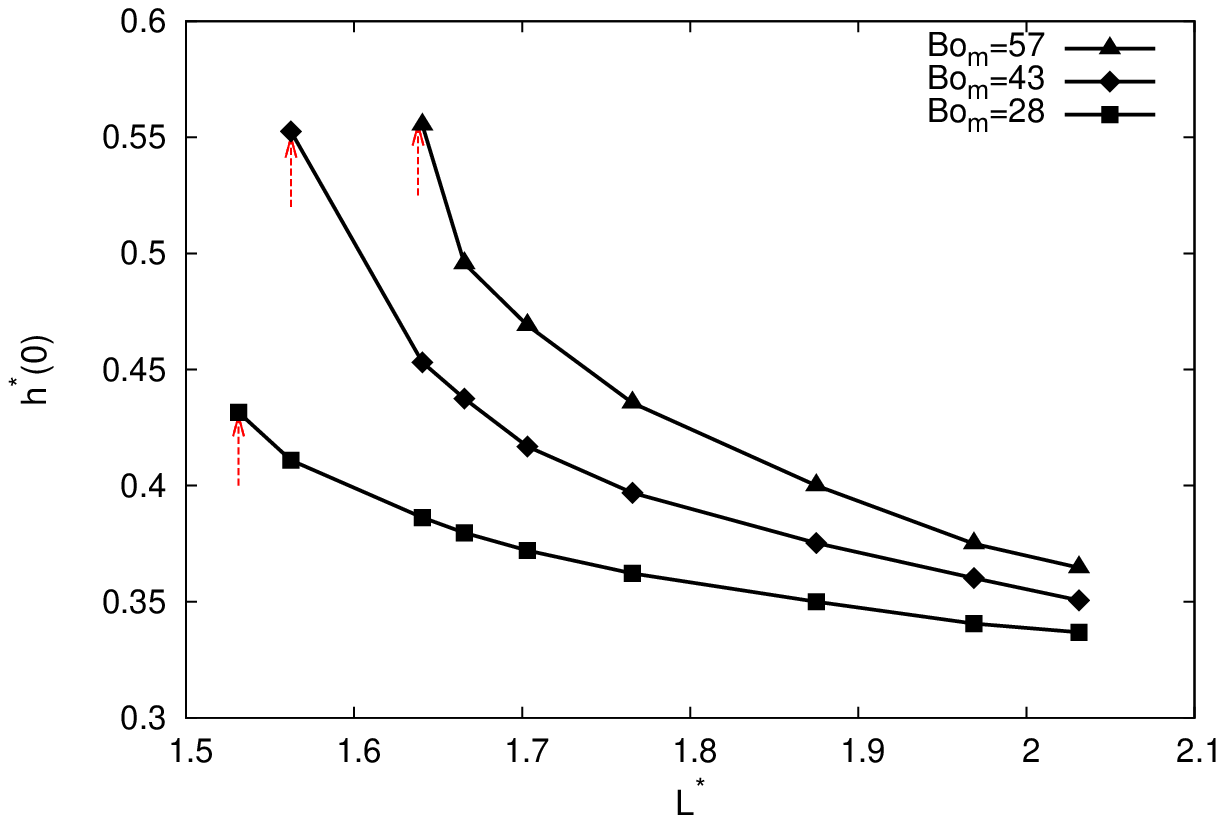}
\end{center} 
\caption{The maximum interfacial deformation, $h^*(0)$, versus the non-dimensional 
         distance between the magnet center and the initially flat film, $L^*$, for various 
         magnetic Bond numbers, ${\mbox{Bo}}_m$. $R_m=3.2$ mm and $\sigma=0.07$ N m$^{-1}$.
         Critical points are marked by red arrows; a smaller $L^*$ will result in no static profiles.}
\label{fig:Bo_m}
\end{figure}
Finally, we investigate the maximum interfacial deformation, $h(0)$, defined as the maximum height of the hump
at $r=0$, as $L$ varies. To characterize this behavior, it is helpful to represent the results
in terms of the following dimensionless parameters \cite{Griffiths13}: the magnetic Bond number
$$
{\mbox{Bo}}_m = \frac{\chi \mu_o M^2_m R_m}{\sigma},
$$
which represents the ratio of the magnetic to surface tension force; and
$$
L^*=\frac{L}{R_m} \quad \mbox{and} \quad h^*=\frac{h}{R_m},
$$
which give the dimensionless distance between the magnet center and the initially undeformed
ferrofluid film; and the dimensionless height of the hump, respectively. In Fig.~\ref{fig:Bo_m},
we present the results of the maximum dimensionless interfacial deformation, $h^*(0)$, predicted by our computational model,
versus dimensionless magnet distance $L^*$, for several values of ${\mbox{Bo}}_m$. 
Again, we see that as the magnet is successively brought closer to the
interface, the maximum hump height increases. We also present the effect of the magnetic Bond number, showing
an increase in the maximum interfacial deformation as the magnetic Bond number is increased.
For all values of ${\mbox{Bo}}_m$ considered, the results show that the maximum interfacial deflection does not vary significantly prior to jetting,
while the magnitude of the maximum deflection increases rapidly for smaller $L^*$. Past the critical values of $L^*$,
no static profile can be obtained (the critical points are shown by red arrows in Fig.~\ref{fig:Bo_m}).
In general, the results indicate that when using a ferrofluid with a smaller susceptibility, the magnet must be placed closer to the interface
to produce jetting. 

\begin{figure}[t]
\begin{center}
 \includegraphics[scale=0.85]{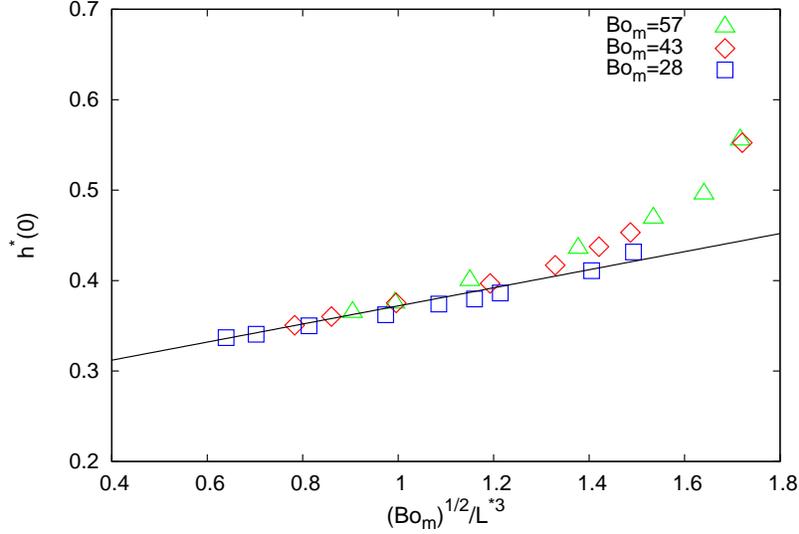}
\end{center} 
\caption{Non-dimensional maximum interfacial deformation, $h^*(0)$, plotted against $({\mbox{Bo}}_m)^{1/2}/L^{*3}$ for 
         the results in Fig.~\ref{fig:Bo_m}. The solid line is the scaling fit, $h^*(0) \propto ({\mbox{Bo}}_m)^{1/2}/L^{*3}$.}
\label{fig:hc}
\end{figure}
Analogous to the study in \cite{Griffiths13}, we identify the natural scaling for the characteristic interfacial 
deflection $h_c$ by
\begin{equation}
 h^2_c=\frac{\mu_o\chi M^2_m R^9_m}{\sigma L^6} = \frac{\mbox{Bo}_m R_m^8}{L^6}, 
 \label{eq:hc}
\end{equation}
representing the dominant balance between
the magnetic and surface tension forces. Numerical results in Fig.~\ref{fig:Bo_m} suggest that, prior
to jetting, the maximum interfacial deflections do not vary significantly for different values of Bo$_m$.
Motivated by this observation, we use Eq.~(\ref{eq:hc}) to determine the scaling relationship between the 
maximum interfacial deflection and the parameters that we vary in our numerical simulations: $L^*$ and ${\mbox{Bo}}_m$. 
In Fig.~\ref{fig:hc}, we plot $h^*(0)$ versus $\sqrt{{\mbox{Bo}}_m}/L^{*3}$, predicted by the scaling model Eq.~(\ref{eq:hc}).
Further from the transition, the simulations results collapse remarkably well onto this single scaling relationship.
Close to the jetting, however, we observe a deviation from the scaling law, whereby the assumption of slowly varying $h^*(0)$ versus $L^*$ breaks down.


\section{Conclusions}
\label{sec:conc}
In this paper, we have carried out direct computations of the deformation of a ferrofluid--air interface under
an external magnetic field gradient generated by a spherical magnet placed at a fixed distance from the interface.
We showed that, for sufficiently large distances, the free surface deformation is static, while below a threshold value,
the interface destabilizes and forms a jetting fluid. These features are also observed experimentally, in a somewhat different system in which the ferrofluid is suspended in water~\cite{Griffiths13}. 
Our numerical results provide data that can be used to determine the maximum deflection of a ferrofluid in the presence of a
magnetic field gradient generated by an external permanent magnet and the consequent transition to jetting. 
Additionally, a simple scaling law allowed us to collapse our numerical results from a series of configurations onto
a single power law.
It is interesting to note that, while the key features observed in the experiments are reproduced in our pure ferrofluid simulations (i.e., static interfacial deformations at large magnet separation 
or low magnetic susceptibility, with transition to jetting at small magnet separations or high magnetic susceptibility), the static interfacial profiles obtained in the two systems are rather different.  
In particular, the interfacial profiles for the water--ferrofluid mix are more ``peaked'' under the same conditions than those for pure ferrofluid; see Fig.~\ref{fig:comparison}. 
This could be due to the water providing a lubricating ``slip'' layer that allows the ferrofluid to be more mobile in the experiments.
Investigating such mixed systems computationally will be deferred to a future publication.

\section*{Acknowledgements}

The authors gratefully acknowledge Howard A. Stone for fruitful discussions and his valuable suggestions.
This research was supported by the National Science Foundation under grants NSF-DMS-1320037 (SA), NSF-DMS-1211713 (LJC), and NSF-DMS-1261596 (LJC). IMG gratefully acknowledges support from the Royal Society through a University Research Fellowship.

\bibliographystyle{elsarticle-num}

\end{document}